\documentclass[11pt,a4paper]{article}   
\usepackage{amsmath}
\usepackage{amssymb}
\usepackage{mathrsfs}

\usepackage{rotating}

\usepackage[round]{natbib}

\newcommand{\real}{\textrm{Re}}

\newcommand{\wrt}{ ~ {\rm d}}

\def\ci{{\mathrm i}}
\renewcommand{\exp}{{\rm e}}

\newcommand{\eg}{e.g.\ }
\newcommand{\ie}{i.e.\ }

\usepackage[normalem]{ulem}
\usepackage{setspace}
\usepackage{xcolor}
\usepackage{marginnote}

\usepackage{xparse}

\NewDocumentCommand{\marcomm}{mo}
 {
  \IfValueTF{#2}
  {\marginnote[\framebox{\parbox{40pt}{\setstretch{1.0}#1}}]{\framebox{\parbox{40pt}{\setstretch{1.0}#1}}}[#2] }
  {\marginnote[\framebox{\parbox{40pt}{\setstretch{1.0}#1}}]{\framebox{\parbox{40pt}{\setstretch{1.0}#1}}}[0pt] }
 }
 
 \NewDocumentCommand{\revmarcomm}{mo}
 {
  \IfValueTF{#2}
  {\reversemarginpar\marginpar[\color{blue}\framebox{\parbox{33pt}{\setstretch{1.0}\scriptsize\sffamily#1}}]{\vspace{#2}\color{blue}\framebox{\parbox{33pt}{\setstretch{1.0}\scriptsize\sffamily#1}}} }
  {\reversemarginpar\marginpar[\color{blue}\framebox{\parbox{33pt}{\setstretch{1.0}\scriptsize\sffamily#1}}]{\vspace{-5pt}\color{blue}\framebox{\parbox{33pt}{\setstretch{1.0}\scriptsize\sffamily#1}}} }
 }
 
\begin{document}
\title{Sea ice floes dissipate the energy of steep ocean waves}
\author{A.~Toffoli$^{1}$, 
L.~G.~Bennetts$^{2}$, 
M.~H..~Meylan$^{3}$, 
C.~Cavaliere$^{3}$, 
A.~Alberello$^{1}$,  
J.~Elsnab$^{4}$,
J. P. Monty$^{4}$ 
\\
{\footnotesize
$^{1}$Centre for Ocean Engineering Science and Technology, Swinburne University of Technology, Hawthorn, VIC, Australia}
\\
{\footnotesize
$^{2}$School of Mathematical Sciences, University of Adelaide, Adelaide, SA, Australia}
\\
{\footnotesize
$^{3}$School of Mathematical and Physical Sciences, University of Newcastle, Callaghan, NSW, Australia}
\\
{\footnotesize
$^{4}$Department of Mechanical Engineering, University of Melbourne, Melbourne, VIC, Australia}
}
\date{\today}
\maketitle

\begin{abstract}
Wave attenuation by ice floes is an important parameter for modelling the Arctic Oceans. At present, attenuation coefficients are extracted from linear models as a function of the incident wave period and floe thickness. Recent explorations in the Antarctic Mixed Ice Zone (MIZ) revealed a further dependence on wave amplitude, suggesting that nonlinear contributions are non-negligible. An experimental model for wave attenuation by a single ice floe in a wave flume is here presented. Observations are compared with linear predictions based on wave scattering. Results indicate that linear models perform well under the effect of gently sloping waves. For more energetic wave fields, however, transmitted wave height is normally over predicted. Deviations from linearity appear to be related to an enhancement of wave dissipation induced by unaccounted wave-ice interaction processes, including the floe over wash.       
\end{abstract}


\section{Introduction}
Waves can penetrate hundreds of kilometres into the ice-covered sea. In doing so, wave energy attenuates exponentially with distance of propagation into sea ice. The rate of decay depends on the wave period and floe thickness \citep[see, for example,][]{wadhams1988attenuation,Squ&Mor80}. Before being completely dissipated, however, waves induce breakup, drift and eventually melt of floes. The interaction between waves and ice contributes to the formation and extent of an interface of scattered ice at the boundary of open waters and pack ice, generally known as the Marginal Ice Zone (MIZ) \citep{wadhams1988attenuation}. The extent of the MIZ plays an
important role in modelling Northern and Southern Oceans and contributes to global climate.


In recent years, the seasonal ice retreat has expanded significantly, leaving large parts of the Arctic Ocean free \citep[e.g.,][]{perovich2008,barber2009}. The emergence of large areas of open water in the summer allows longer fetches for wave generation. As a result, increasing significant wave height  approaches the ice edge \citep{francis2011,asplin2012,thomson2014swell}. This contributes more substantially to the dynamics of the MIZ \citep[cf.][]{kohout2014} than ever before. 

The majority of waves-in-ice modelling investigations focus on predicting the rate of wave attenuation. 
The ratio of the prevailing  floe diameters to the incident wavelengths determines the mechanisms responsible for reducing wave energy. In a  field of  foes with diameters much smaller than their wavelength (\eg pancake ice), wave energy is reduced with penetration distance due to viscous losses \citep[e.g.,][]{wang2011continuum}. In contrast, floes with diameters comparable to or larger than the wavelength (\eg floes produced by wave-induced breakup) reflect a proportion of the incident wave energy, dissipate a proportion and transmit the remaining proportion. Under these circumstances, models based on the elastic bending of the floes can be used to approximate the underlying physics \citep[see, for example,][among others]{kohout2008elastic,bennetts2010wave}. Models, however, only consider the interaction of the waves with an elastic boundary layer at the surface of the water. For simplicity, only small-amplitude waves are considered. Furthermore, other effects such as viscosity, floe collisions, submergence of the floe and nonlinear effects are neglected. As a consequence, transmission is simply defined by the proportion of reflected energy, while other sources of dissipation are neglected.  

Scattering models have been recently coupled with the spectral wave models to simulate waves in Antarctic MIZ \citep{Dob&Bid13}. Despite limitations, hindcasts are overall consistent with buoy data. The model, however, seems to mispredict significant wave height at the peak of the storm \citep[see, for example, figure 4 in][]{Dob&Bid13}. In more severe conditions, the floe is subjected to water running over the top of it (overwash), drift, six rigid-body degrees of freedom (\ie heave, surge, sway, pitch roll and yaw), besides elastic motions. 
These processes enhance wave dissipation and their extent is proportional to wave amplitude. Interestingly enough, storm waves recorded in the Antarctic MIZ \citep{Meyetal14} highlighted how the attenuation rate depends indeed on both wave period and wave amplitude (or, in other word, the wave steepness, a proportion of the ratio of height to wavelength). It is therefore conjectured that current wave-ice models may misjudge attenuation coefficient, especially for large waves. In the present letter, an experimental model of wave transmission by an single ice floe is discussed to assess the extent of the departure of the underlying physics from model predictions. 


\section{Experimental model} \label{exp}

Transmission of an incident wave field by an artificial ice floe was monitored in the Extreme Wind-Wave Flume of the University of Melbourne, Australia. The flume is 60\,m long and 1.8\,m wide. Water depth was set to $H=0.8$\,m. The facility is equipped with a cylinder type wave-maker on one end for the mechanical generation of waves. 
A linear beach with slope 1:10 is deployed at the opposite end to ensure wave absorption. 
A detailed analysis of the beach performance revealed a contemned reflected waves with energy content up to 10\% the energy of the input component. 

\begin{figure}
 \centering
\includegraphics[width=0.8\textwidth]{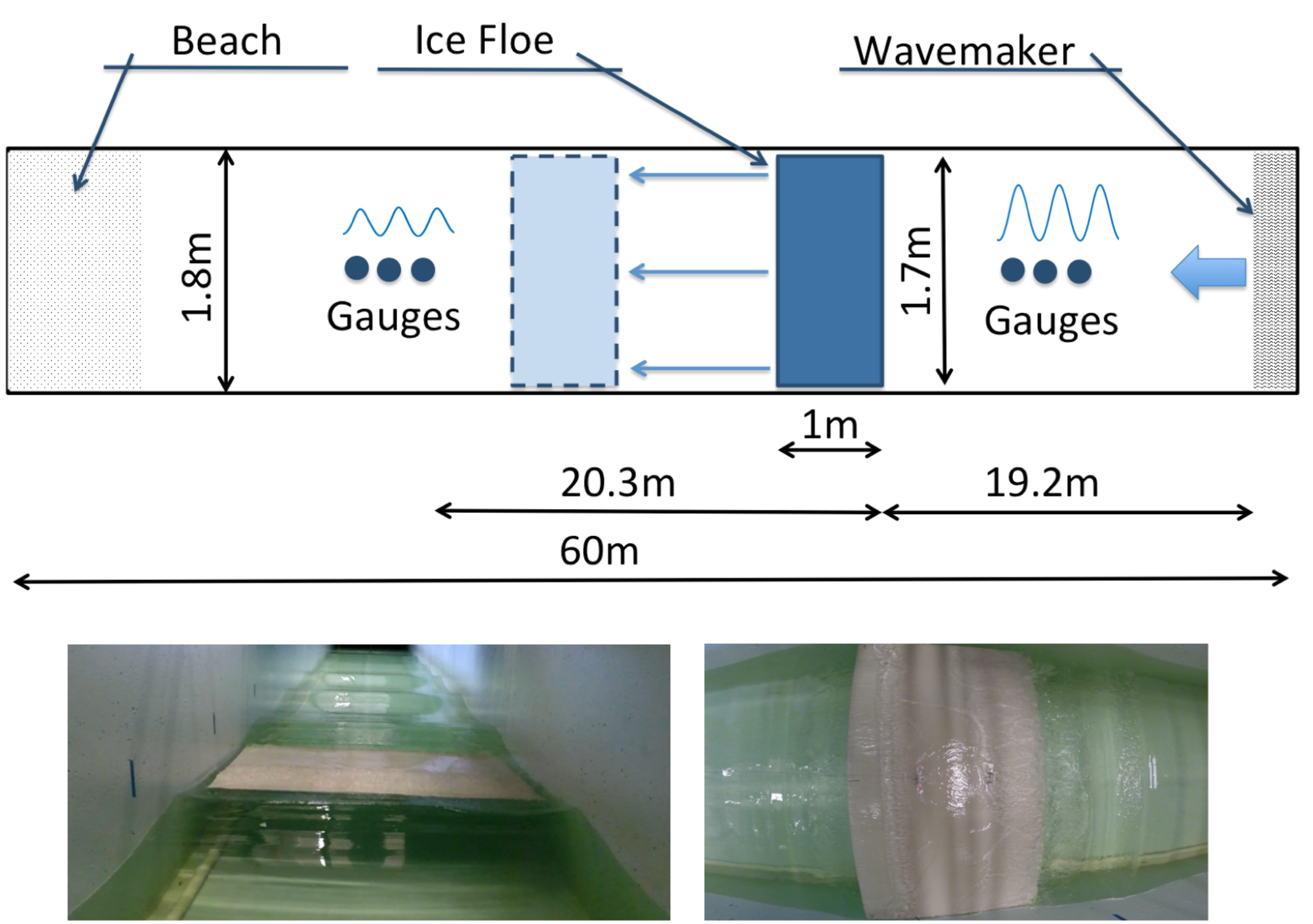}
 \caption{Schematic of the experimental setup (not to scale). The model floe was represented by a polypropylene sheet 1.7\,m wide and 1\,m long. No mooring was applied. Waves were generated by a cylindrical wave maker at one (right-hand) side and absorbed by a linear (1:10) beach at the other (left-hand) side. Water surface elevation was recorded by six capacitance gauges: three in front and three in the rear of the floe. Front and rear probes were deployed approximately 20.3 $m$ apart to ensure enough space for the model floe to drift under the influence of waves. Photos in the bottom panels highlight floe's behaviour in water. Bottom right panel, particularly, show water running on top of the floe (over wash).}
 \label{setup}
 \end{figure}

A model ice floe was represented by a $h=10$\,mm thick polypropylene plastic plate and cut into a rectangular shape 1\,m long and 1.7\,m wide. The Young's Modulus of polypropylene is 1600\,MPa and its density is 0.905\,g\:cm$^{-3}$. The plate was deployed at 19.2\,m from the wave-maker. No mooring was applied to the floe. The free-floating configuration allows drift, overwash, the six rigid-body degrees of freedom and elastic motions to occur simultaneously. A schematic of the experimental setup is presented in Fig. \ref{setup}.  

Incident regular waves were obtained by imposing a wave period and amplitude at the wave-maker. 
The wave-maker only produces plain sinusoidal waves. Bound modes typical of Stokes waves appear automatically soon after wave generation. Four periods were applied: $\tau= 0.7$\,s, 0.8\,s, 0.9\,s and 1\,s. 
These correspond to wavelengths $\lambda$ of 0.76\,m, 1\,m, 1.26\,m and 1.56\,m, respectively. 
For each period, six nominal amplitudes $a$ were imposed. This defines six different values of wave steepness $ka$, where  $k=2\pi/\lambda$ is the wavenumber associated to the wave period. The following values were considered: $ka=0.04$, 0.06, 0.08, 0.1, 0.12 and 0.14, ranging from gently sloping to storm-like waves. Note that wave steepness is a measure of wave nonlinearity and hence it defines the intensity of bound modes (\ie second-order-like nonlinearity).

The water surface elevation was monitored by six capacitance wave gauges at a sampling frequency of 1000\,Hz. 
Three of them were deployed in front of the incident edge of the floe to capture reflected wave components. Other three gauges were deployed at a distance of 20.3\,m from the incident edge (see Fig. \ref{setup}) to measure the transmitted wave field. This distance ensures enough space for the model floe to drift under the influence of waves. From visual observations (based on camera recording and markers at the wall), the floe propagated at a speed between 10 and 50 times greater than the Stokes drift, depending on wave amplitude and period. 

The floe is overwashed by waves. Records of the overwashed water were gathered with two small probes deployed on the top of the floe in a post-experiment phase. For technical reason related to the length of wires, the floe was moored during overwash measurements. Overall, overwashed layer was observed to range from a few millimetres for gently sloping waves to a few tens of millimetres for the storm-like waves. It is worth mentioning that the overwash is generated at the front and rear ends of the floes, alternately as they pitched. Steep, shallow-water waves propagated in the overwashed fluid itself 
(see bottom left photo in Fig. \ref{setup}). When shallow-water waves, travelling in opposite directions, up and down the floes, meet, waves often become large enough to break, which dissipates wave energy. An example of floe propagation in steep incident waves ($\tau=1$\,s and $ka=0.14$) and concurrent breaking is shown in the supplementary material.

For each run, 60\,s time series of surface elevation were produced. Such a short period of time excluded any contamination of reflected waves in the measurement area. Three repetitions of the same test were carried out to estimate experimental uncertainties. As the nominal amplitude may differ from the one obtained in the flume, the incident wave was recorded at the first 3 gauges in the absence of the model floe for benchmarking.  

  
\section{Theoretical model} \label{numerics}

Consider a two-dimensional model of a regular incident wave interacting with a floe, 
with a horizontal dimensional and a depth dimension. 
Let the Cartesian coordinate $(x,z)$ denote locations in the water.
The coordinate $x$ defines the horizontal location. It points in the direction of the incident wave and has its origin set to coincide with the front edge of the floe at rest.
The coordinate $z$ defines the vertical location. It points upwards and has its origin set to coincide with the water surface at rest.

Linear potential-flow theory is used to model water motions. 
The water is, therefore, assumed to be homogeneous, inviscid, incompressible and in irrotational motion.
It follows that the water velocity field can be defined as the gradient of a scalar velocity potential, denoted $\Phi(x,z,t)$, 
where $t$ is time.

Kirchhoff-Love thin-plate theory is used to model the floe, which covers the water surface over the interval $x\in(0,l)$.
The floe, therefore, bends in response to the wave motion, 
in addition to responding in its rigid-body motions.
The rigid motions in the vertical direction consist of translational heave and rotational pitch.
Further, the floe surges back and forth in the $x$-direction.
Thin-plate theory permits the deformation of the floe to be defined in terms of the vertical displacements of its lower surface, 
denoted $z=-d+w(x,t)$, where $z=-d$ is the Archimedean draught of the floe.
Surge is defined by the horizontal location of the floes centre of mass, denoted $u(t)$.

Wave amplitudes are assumed to be sufficiently small that linear theory is valid.
The water-floe system therefore oscillates at the frequency of the incident wave.
The velocity potential, displacement function and surge can, therefore, be expressed as 
$\Phi(x,z,t)=\real\{(g/\ci\omega)\phi(x,z)\exp^{-\ci\omega t}\}$, 
$w(x,t)=\real\{\zeta(x)\exp^{-\ci\omega t}\}$
and $u(t)=\real\{\xi\exp^{-\ci\omega t}\}$, where $\omega=2\pi/\tau$ is angular frequency and $\phi$, $\zeta$ and $\xi$ are complex-valued.

The (reduced) velocity potential, $\phi$, satisfies Laplace's equation in the water domain, an impermeable floor condition
and the linearised free-surface condition at points not covered by the floe, \ie
\begin{equation}
\phi_{xx}+\phi_{zz}=0
\quad
\text{for}
\quad
(x,z)\in\Omega,
\quad
\phi_{z}=0
\quad
\text{for}
\quad
z=-H,
\quad
\phi_{z}=\sigma\phi
\quad
\text{for}
\quad
x\notin(0,l)
\;\;
\text{and}
\;\;
z=0,
\end{equation}
respectively, where $\sigma=\omega^{2}/g$ is a frequency parameter.
The water and floe motions are coupled at by dynamic and kinematic conditions applied at the wetted surface of the floe at rest.
The conditions in the horizontal direction are
\begin{equation}
\phi_{x}=\sigma\xi
\quad
\text{for}
\quad
x=0,l\;\;
\text{and}\;\;
z\in(-d,0),
\qquad
\text{and}
\qquad
-\sigma
hl
\xi
=
\int_{-d}^{0}
[\phi]_{x=0}^{l}
\wrt z
.
\end{equation}
The conditions in the vertical direction are
\begin{equation}
\phi_{z}
=
\sigma \zeta
\qquad
\text{and}
\qquad
(1-\sigma d)\zeta
+F\zeta''''
=
\phi
\quad
\text{for}
\quad
x\in(0,l)\;\;
\text{and}\;\;
z=-d,
\end{equation}
where $F=Eh^{3}/\{12\rho g(1-\nu^{2})\}$ is a scaled flexural rigidity of the floe and $\nu=0.3$ is Poisson's ratio.

On the incident wave side of the floe, far enough away from the floe that the exponentially decaying local motions have died out, the wave field is the sum of the incident, $\phi_{I}$ wave plus a reflected wave, $\phi_{R}$.
On the opposite side of the floe, far enough away from the floe, the wave field is composed of a transmitted wave only, $\phi_{T}$.
The incident, reflected and transmitted wave potentials are defined as
\begin{equation}
\phi_{I}
=
\frac{a\exp^{\ci k x}\cosh k(z+H)}{\cosh kH}
,
\quad
\phi_{R}
=
\frac{R\exp^{-\ci k x}\cosh k(z+H)}{\cosh kH}
\quad
\text{and}
\quad
\phi_{T}
=
\frac{T\exp^{\ci k x}\cosh k(z+H)}{\cosh kH}
,
\end{equation}
where $R$ and $T$ are the reflected and transmitted amplitudes, which must be obtained as part of the solution.
\citet{Mey&Squ94} devised the first method to the solve the above problem, without surge, $\xi=0$ 
and using the shallow-draught approximation, $d=0$. 
\citet{Ben&Chu11} developed a solution method for the problem with surge and non-zero draught.
The model predicts the proportions of incident wave energy reflected and transmitted to be $\mathcal{R}=\vert R/a\vert^{2}$ and $\mathcal{T}=\vert T/a\vert^{2}$, respectively.
The model is conservative, \ie no energy loss, and thus $\mathcal{R}+\mathcal{T}=1$.

\section{Results} \label{surface}

Transmitted wave heights were calculated by post-processing records in the lee of the floe with a standard zero-crossing analysis \citep[e.g.,][]{emery01}. Time series were first low- and high-pass filtered to remove contaminating components greater than 5.5 times and smaller than 0.35 times the dominant frequency. Individual, transmitted wave heights were then extracted at each of the three gauges and for each of the three realisations. Both down-crossing and up-crossing height were considered. To generalise the results, the steepness of individual waves (\ie $ka$, where the amplitude $a$ is defined as half the individual wave height $H$) is considered herein. 

 \begin{figure}
 \centering\includegraphics[width=0.8\textwidth]{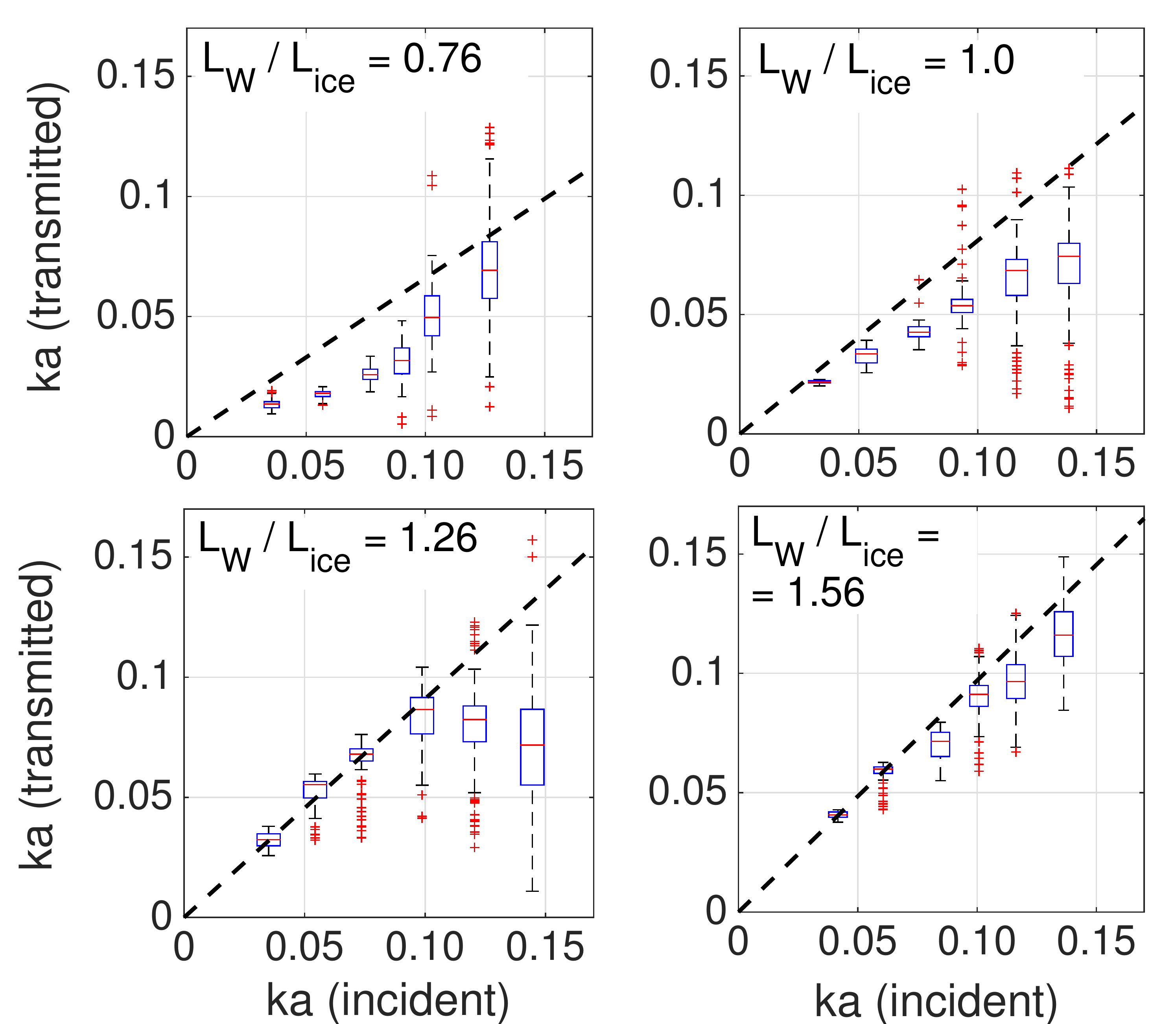} 
 \caption{Transmitted versus incident wave steepness: experimental data (box-and-whiskers); model predictions (dashed lines). Box-and-whiskers includes all individual transmitted wave steepness for the concurrent incident steepness. For each box, the central mark represents the median, while the edges of the box are the $25^{th}$ and $75^{th}$ percentiles. The whiskers extend to the most extreme data points, which are not considered outliers; outliers are plotted individually (+ symbols).}
 \label{tr_mooring}
 \end{figure}

Scatter plots of transmitted versus incident wave steepness are presented in Fig.\ \ref{tr_mooring} for the different incident wavelengths. A box-and-whisker representation is applied to better express the variability of the transmitted filed. The tops and bottoms of each box represents the $25^{\rm th}$ and $75^{\rm th}$ percentiles of the samples, respectively. The middle line is the sample median. Whiskers extend to the most extreme values. Observations beyond the whiskers (+ symbols in the figures) are considered outliers. Data uncertainty is found to be clearly affected by the steepness. The largest variability was observed for high steep waves, where breaking was most likely to occur.  Numerical predictions of transmitted steepness, as a function of the incident steepness, are reported too (see dashed line in Fig.\ \ref{tr_mooring}). We remark that model predictions only depends on the wave period/length. 

For the longest wavelength (i.e. $\lambda / l = 1.56$), the floe exerts a minimal effect on wave transmission, in agreement with field observations \citep[cf.][]{wadhams1988attenuation}. Overall, experimental observations fit model data within the range of experimental uncertainty. The median values (middle red line in the box), nevertheless, suggest a weak departure from model predictions, with transmitted wave height being slightly overpredicted for steep conditions. Under these circumstances, shallow-water waves on the overwashed layer becomes very steep and break, increasing the level of dissipation. 
Departure from model predictions becomes more pronounced with decreasing wavelengths. 
For  $\lambda / l  = 1.26$, the model performs well up to an incident $ka\approx 0.1$. 
Transmission of steeper waves, however, is substantially overpredicted by the models; departure is well outside the most extreme values limits indicated by the whiskers. This highlights a substantial dependence on wave amplitude, in agreement with records in Antarctic MIZ \citep{Meyetal14}. The model performs least well for $\lambda / l  \leq 1.0$, where deviations are recorded over the entire range of incident steepness.  For the shortest waves ($\lambda / l = 0.76$\,m), deviations appears to be less significant for incident $ka>0.1$.    

\begin{figure}
 \centering
 \includegraphics[width=0.6\textwidth]{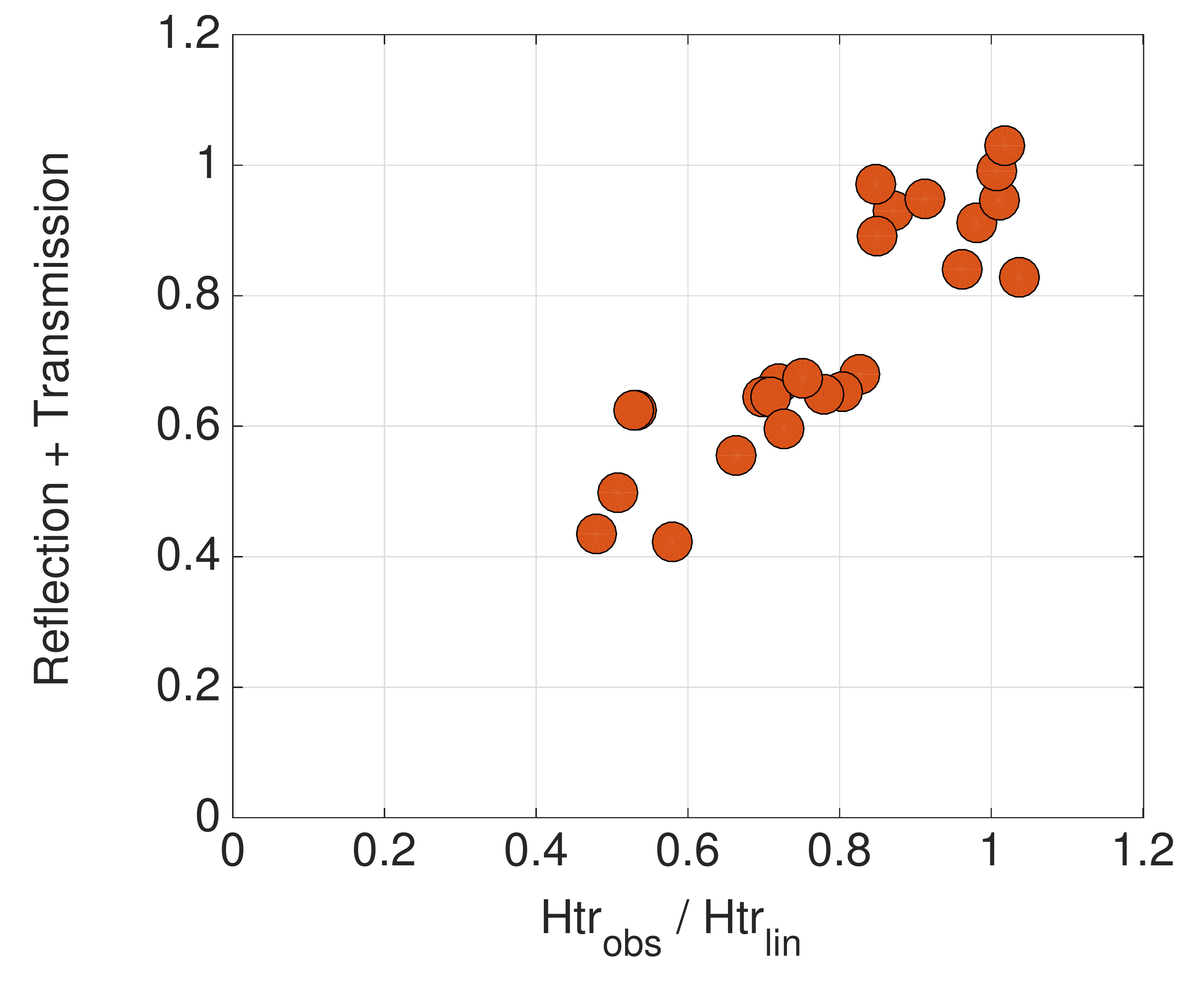} 
 \caption{Sum of reflected and transmitted energy versus deviation from linear prediction.}
 \label{comparison}
 \end{figure}

We remark that the wave-ice model relies on a balance between reflection and transmission and hence other sources of wave dissipation are neglected. To verify that deviations takes place when the balance is violated, the sum of a representative reflection and transmission coefficient is presented as a function of the extent of deviations from model prediction in Fig.\ \ref{comparison}. Reflection and transmission coefficients are calculated as the change in spectral variance ($m0$) with respect to the incident conditions in front and in the rear of the floe, respectively:
\begin{equation}
\mathcal{R}=\frac{|m0_{front}-m0_{inc}|}{m0_{inc}} ; \: \mathcal{T}=1-\frac{|m0_{rear}-m0_{inc}|}{m0_{inc}}.
\end{equation}   
A representative spectrum for each test is obtained by first calculating spectra at consecutive windows of 4096 points (with no overlap) and then averaging them over the entire time series. Deviations from model prediction are estimated as the ratio of median transmitted wave height to the incident wave height ($\beta$) for each tests. 

When the extent of transmission coincides with the amount of reflected energy ($\mathcal{R}+\mathcal{T}=1$), the model performs well and hence no significant deviations are observed ($\beta \approx 1$). Primarily for steep waves, however, this balance is violated (\ie $\mathcal{R}+ \mathcal{T}  < 1$) because of, but necessarily limited to, wave breaking on the overwashed fluid. As additional sources of dissipation becomes more relevant, the model under performs.

\section{Conclusions}

Experiments in the Extreme Wind-Wave Flume at the University of Melbourne, Australia, were conducted to measure wave transmission by a single model ice floe. Observations were compared with model predictions to evaluate the extent of model assumptions. The ice floe  was modelled with an unmoored 10\,mm polypropylene plate. Water surface elevation was monitored in front and in the rear of the floe. Tests were carried out with regular  waves with different periods and  amplitudes, ranging from gently sloping to storm like waves. 

Analysis of the experimental data indicates that models perform generally well under the effect of gently sloping waves. For steep (storm-like) wave conditions, additional source of wave dissipation like wave breaking on top of overrated fluid affect wave transmission substantially. As breaking dissipation violate the assumed balance between reflection and transmission, the model substantially under perform.   



\section*{Acknowledgements}

This work was funded by the Australian Research Council.


\end{document}